\def\spose#1{\hbox to 0pt{#1\hss}}

\def\multleft#1{\hbox to size{\vbox {\halign {\lft{##}\cr #1}}\hfill}\par}
\def\multright#1{\hbox to size{\vbox {\halign {\rt{##}\cr #1}}\hfill}\par}

\def\today{\ifcase\month\or January\or February\or March\or April\or May\or
      June\or July\or August\or September\or October\or November\or December\fi
      \space\number\day, \number\year}
\def\$<${\thinspace}
\def\s{\hbox{\phantom{5}}}	%one space
\def\cm{{\rm\thinspace cm}}

\def\erg{{\rm\thinspace erg}}

\def\s{{\rm\thinspace s}}

\def\ergpcmsqps{\hbox{$\erg\cm^{-2}\s^{-1}\,$}}

\def\H2{\hbox{H$_{2}$}}

\documentstyle[psfig]{mn}

\begin{document}
\hsize=6truein

\title{A survey of molecular hydrogen in the central galaxies of cooling flows}
\author[A.C. Edge et al.]
{\parbox[]{6.in} {A.C.~Edge$^1$, R.J.~Wilman$^{2,3}$, R.M.~Johnstone$^2$, C.S.~Crawford$^2$, A.C.~Fabian$^2$ and S.W.~Allen$^2$\\ \\
\footnotesize
1. Department of Physics, University of Durham, Durham, DH1 7LE. \\
2. Institute of Astronomy, Madingley Road, Cambridge CB3 0HA. \\ 
3. Leiden Observatory, P.O. Box 9513, 2300 RA Leiden, The Netherlands.}} 

\maketitle

\begin{abstract}
We present a large sample of H- and K-band spectra of 32 optically 
line-luminous central cluster galaxies. We find significant rovibrational 
\H2 emission in 23 of these galaxies as well as H recombination and/or
[FeII] emission in another 5. This represents a fourfold increase
in the number of molecular line detections known.
A number of the detections are of extended emission (5--20~kpc).
In several objects we find significant [SiVI] emission that appears to 
correlate with the strength of high ionization lines in the optical (e.g. [OIII]). 
This comprehensive sample builds on previous work and
confirms that warm (1000--2500~K) molecular
hydrogen is present wherever there is ionized material in the cores
of cooling flows and in most cases it also coincides with CO molecular
line emission. 

\end{abstract}


\section{INTRODUCTION}

The cooling time of hot X-ray-emitting gas in the central regions of
massive, relaxed clusters of galaxies can be substantially less than
the Hubble time.  The gas in these regions can thus cool and recombine,
initiating a cooling flow (Fabian \& Nulsen 1977; Cowie \& Binney
1977).  The ultimate fate of this cooling gas has been the subject of
an extensive and strongly contested debate (see Fabian 1994). The cold gas
has not been detected in molecular form and so is inferred to reside in a
phase with $T_{\rm gas} << 100$\,{\sc k}. Calculations of the gas
properties are consistent with current observed limits (Ferland, Fabian
\& Johnstone 1994).  

Until recently, the only cooling flow known to contain 
molecular gas was that around NGC~1275 in Perseus (see Bridges \& Irwin 1998), 
although the interpretation of this source is
complicated by the strongly varying nuclear component.  Moreover, the
presence of the molecular gas may be related to the apparently on-going merger in
this system, which has been the subject of a long-running debate (Van
den Bergh 1977; Hu et al.\ 1983; Pedlar et al.\ 1990; Holtzman et
al.\ 1992; Norgaard-Nielsen et al.\ 1993).  Another observational window 
on molecular hydrogen lies in the near-infrared where a
number of strong rovibrational lines fall in the K-band. These
lines have been detected in 8 different cooling flow central cluster
galaxies (Jaffe \& Bremer 1997; Falcke et al.\ 1998; Krabbe et al. 2000;
Wilman et al.\ 2000;
and Jaffe, Bremer \& van der Werf 2001). The detection of
molecular gas at 1000-2500~K implies excitation
is required and that only a very small amount of `hot' molecular
gas is present. So there is evidence for some molecular
gas in the cores of cooling flows but its relation to the
deposited gas predicted from the X-ray observations is
far from clear.

This situation has
been radically altered with the detection of CO line emission in
16 central cluster galaxies by Edge (2001). The molecular
gas masses of 10$^{9-11.5}$ M$_\odot$ imply that around 2--10\%
of the total deposited gas predicted from previous X-ray observations
can be accounted for directly. 
However, with the dramatically reduced mass deposition
rates found in {\it Chandra} and {\it XMM-Newton} observations
(Allen, Ettori \& Fabian 2001, Schmidt, Allen \& Fabian 2001, Peterson et al. 2001),
the molecular gas masses derived from CO observations are within
a factor of three of that now expected.
The mass of molecular gas appears to correlate best
with the optical emission line luminosity indicating that
the molecular gas observed is warmed from a  lower temperature.
The rapid progress in this field has been possible
due to the selection of massive cooling flows from the
ROSAT All-Sky Survey (see  Crawford et al. 1999), few
of which were known before 1990. To provide a direct
comparison of the `hot' (1000--2500~K) and `cool' (30~K)
molecular gas components requires near-infrared
spectra of the systems that have been searched for CO.

In this paper we present the spectra and analysis for a systematic
search for near-infrared hydrogen molecular lines in a complete
sample of strong optical emission line emitting central galaxies 
drawn from Crawford et al. (1999). 
An ${H\alpha}$ flux limit of $>3\times10^{-15}$~erg~cm$^{-2}$s$^{-1}$
was adopted to ensure moderately bright lines and a redshift range of
0.03  to 0.3 to cover at least one \H2 line in the K-band. This
selection produces a sample of 18 objects reachable with UKIRT 
(one, A2146, is above the declination limit of 60$^\circ$).
Of this sample, only one object, A478, has not been observed in this study
but it has a published spectrum
(Jaffe, Bremer \& van def Werf 2001). We also include seven weaker
line emitting or lower redshift objects from Crawford et al. (1999)
to fully sample any potential range
in recombination lines to molecular line ratios,
seven non-BCS central cluster galaxies
with strong optical emission lines drawn from the literature
(e.g. NGC~1275, A2597 and PKS~0745$-$191), 
and one control object with no lines (A2029) giving a total sample
of 32 spectra. We also extend our spectral
coverage to [FeII] in the majority of our targets making this the
first comprehensive study of this line in central cluster galaxies.

Throughout we assume $\Omega_0=1$ and $H_0 = 50$\,km\,s$^{-1}$\,Mpc$^{-1}$.

\section{OBSERVATIONS}

The observations presented in this paper were taken with the
CGS4 spectrograph on the United Kingdom Infrared Telescope (UKIRT)
in September 1999 and March 2000, as shown in the observing log
in Table~1. The 256$\times$256 InSb array, 40 l/mm grating and
300 mm focal length camera were used giving a spatial scale
of 0.61 arcsec per pixel. With a 2-pixel (1.22$''$) wide slit this set-up
achieves spectral resolutions of
880 and 570 km~s$^{-1}$ FWHM in the H- and K-bands respectively.
In all but two observations, the slit was oriented at a position
angle of 0 degrees (north-south).
The NDSTARE mode was used along with the conventional object-sky-sky-object
nodding pattern. Atmospheric absorption features were removed by
ratioing with main sequence F stars and the spectra were calibrated
against photometric standards. We add in one short, archival observation
(A262, a CO detection made after our March 2000 run) which used the
75 l/mm grating with a 1-pixel (0.61$''$) slit at a position angle of 2.9$^\circ$.
The spectra were reduced using version
V1.3-0 of the Portable CGS4 Data Reduction package available through 
Starlink. 

\begin{table*}
\caption{Log of UKIRT Observations. The seeing in all cases was between 0.8--1.2$''$ and the position angle 
was 0~deg unless stated otherwise.}
\begin{tabular}{lcccc}
%\noalign{\medskip \hrule \medskip}
Cluster   & Redshift  &  Date & wavelength  & exposure \\
          & &       & ($\mu$m)    & time (min) \\ \hline
 & & &   \\
Zw8276    & 0.075 \ & 02/09/99 & 1.87--2.49  &  64 \\
A2390     & 0.2328 & 02/09/99 & 1.87--2.49  & 80 \\
Zw235     & 0.083 \ & 02/09/99 & 1.87--2.49  &  64 \\
RXJ0352+19 & 0.109 \ & 02/09/99 & 1.87--2.49  & 48 \\
RXJ0338+09 & 0.0338 & 02/09/99 & 1.87--2.49  & 80 \\
 & & &   \\
Zw8193    & 0.1754 & 03/09/99 & 1.87--2.49  & 64 \\
Zw8197    & 0.1140 & 03/09/99 & 1.87--2.49  & 40 \\
A2597     & 0.0852 & 03/09/99 & 1.87--2.49  & 64 \\
A11       & 0.151 \ & 03/09/99 & 1.87--2.49  & 40 \\
A85       & 0.0555 & 03/09/99 & 1.87--2.49  & 40 \\
RXJ0439+05 & 0.208 \ & 03/09/99 & 1.87--2.49  & 32 \\
NGC1275   & 0.0176 & 03/09/99 & 1.87--2.49   & 24 \\
 & & &   \\
RXJ0821+07 & 0.110 \ & 21/03/00 & 1.87--2.49  & 84 \\
A795      & 0.1355 & 21/03/00 & 1.87--2.49  & 60 \\
A1068     & 0.1386 & 21/03/00 & 1.87--2.49  & 80 \\
A1664     & 0.1276 & 21/03/00 & 1.87--2.49  & 48 \\
A1835     & 0.2523 & 21/03/00 & 1.87--2.49  & 60 \\
A2204     & 0.1514 & 21/03/00 & 1.87--2.49  & 52 \\
 & & &   \\
A646      & 0.1268 & 22/03/00 & 1.87--2.49  &  80 \\
4C+55.16  & 0.242 \ &  22/03/00 & 1.87--2.49  &  40 \\
A1795     & 0.062 \ & 22/03/00 & 1.87--2.49  &   40 \\
Her-A     & 0.154 \ & 22/03/00 & 1.87--2.49  &   41 \\
 & & &  \\
RX0747-19 & 0.1028 & 23/03/00 & 1.87--2.49  &  80 \\
Zw3146    & 0.2906 & 23/03/00 & 1.87--2.49  &  60 \\
Zw3916    & 0.204 \ & 23/03/00 & 1.87--2.49  &   30 \\
A1885     & 0.090 \ & 23/03/00 & 1.87--2.49  &   44 \\
A2029     & 0.0786 & 23/03/00 & 1.87--2.49  &  52 \\
A2199     & 0.031 \ & 23/03/00 & 1.87--2.49  &   52 \\
 & & &   \\
RX0821+07 & 0.110 \ & 24/03/00 & 1.87--2.49  &   16 (PA=77 deg) \\
RX0821+07-off & 0.110 \ & 24/03/00 & 1.87--2.49  & 44 (PA=77 deg) \\
Hydra-A   & 0.0538 & 24/03/00 & 1.87--2.49  & 40 \\
Zw7160    & 0.2578 & 24/03/00 & 1.87--2.49   & 48 \\
A2052     & 0.0351 & 24/03/00 & 1.87--2.49   & 52 \\
 & & &   \\
RX0747-19 & 0.1028 & 25/03/00 & 1.35--1.96  &  32 \\
RX0821+07 & 0.110 \ & 25/03/00 & 1.35--1.96  &  12 (PA=77 deg) \\
RX0821+07-off & 0.110 \  & 25/03/00 & 1.35--1.96  & 28 (PA=77 deg) \\
A1068     & 0.1386 & 25/03/00 & 1.35--1.96  &  28 \\
Hydra-A   & 0.0538 & 25/03/00 & 1.35--1.96  & 24 \\
Zw3146    & 0.2906 & 25/03/00 & 1.35--1.96  &  42 \\
A1664     & 0.1276 & 25/03/00 & 1.35--1.96  &  32 \\
A1795     & 0.062 \ & 25/03/00 & 1.35--1.96  &   40 \\
A2052     & 0.0351 & 25/03/00 & 1.35--1.96  &  40 \\
A2204     & 0.1514 & 25/03/00 & 1.35--1.96  &  60 \\
 & & &   \\
A262      & 0.0166 & 12/08/97 & 1.82-2.48 & 2 \\
\noalign{\smallskip \hrule}
\end{tabular}
\end{table*}

\section{RESULTS}

The individual K- and H- band spectra are presented in Figures 1 and 2 respectively
in the order they appear in Table~1. The line fluxes are summarised in Table~2
in RA order.

Several of the observations show significant extent or velocity structure
or were selected from the literature,
so we discuss these individually here:

\noindent {\bf A11} Selected as X-ray luminous radio galaxy by Perlman et al.\ (1998)
but it is a central cluster galaxy. This galaxy was included in this study
on the basis of a published spectrum and not
on X-ray flux (which is below 3$\times 10^{-12}$ erg cm$^{-2}$ s$^{-1}$ 0.1--2.4~keV).

\noindent {\bf RXJ0747-19 aka PKS~0745-191} This system has {\it HST} narrow band imaging with
{\it NICMOS} (Donahue et al.\ 2000) that shows extended emission. Our CGS4
spectrum does not sample the E-W extension highlighted by Donahue et al.\ (2000)
but shows significant extent in both Pa$\alpha$ and 1-0 S(1) and S(3) over
5 pixels (3.0$''$) or 8~kpc to the North, consistent with the {\it HST} data.
See section 4.4 for further discussion.

\noindent {\bf RXJ0821+07} This system is highly anomalous in virtually
every waveband so far observed and the NIR is no exception. A strong,
velocity-offset CO detection is made in this cluster (Edge 2001) and
the optical morphology of the central galaxy is very peculiar in
a recent HST snapshot with an infalling galaxy apparently being disrupted
(Bayer-Kim et al.\ 2002).
We made an additional CGS4 observation to cover the region of this
infalling galaxy but made no significant detection of any
H$_2$ lines although Pa$\alpha$ was extended over 4 pixels (2.4$''$) or 6~kpc to
the North. The lack of NIR H$_2$ emission is surprising given
the strength of the CO line. This could be an indication that the
mechanism exciting the molecular gas in RXJ0821+07 is different
from that behind the majority of the other detections presented here.

% refer to Bayer et al 2002?

\noindent {\bf 4C+55.16} The central galaxy in this cluster
contains a very strong flat spectrum radio source so it is
possible that the S(3) line is blended with [SiVI]. At the
resolution used in this work it is not possible to determine
the contribution of each line. This cluster was included in this
study on the basis of its known cooling flow (Iwasawa et al. 1999)
and does not fall above the X-ray flux limit of the BCS and hence
not in the Crawford et al. (1999) sample.

\noindent {\bf A1068} This system is one of the strongest optical
line emitters in the sample. Both the Pa$\alpha$ and H$_2$ 
lines are extended to the south by 4 pixels (2.4$''$) or 8~kpc
(see section 4.4).

\noindent {\bf A2204} The Pa$\alpha$ and H$_2$ 
lines are also extended to the south by 5 pixels
(3.0$''$) in this system but show also
a velocity shift of $+$200--300~km~s$^{-1}$. The [FeII]
for this source is particularly strong (see Section 4.2).

\noindent {\bf Zw8193} This system is the most complex spatially
and in velocity. From the original discovery spectrum in Allen et al.
(1992) it was noted that the absorption  and emission lines
were not coincident in velocity. The reasons behind this are 
evident from our CGS4 spectra which showed two distinct components
3$''$ or 12~kpc apart (but both contained in the INT FOS aperture) that
each show lines but with a velocity difference of 400--500~km~s$^{-1}$.
There is extended line emission between the two components (see section 4.4).
This cluster is a prime candidate for observation with the
next generation of infrared integral field units.

\noindent {\bf A2597} This galaxy has {\it HST} narrow band imaging with
{\it NICMOS} (Donahue et al.\ 2000) and, as in the case of RXJ0747-19,
the extended emission over 7 pixels (4.2$''$ or 9~kpc) we find to the North
is consistent with these higher spatial resolution {\it HST} data.

\noindent {\bf NGC1275} This galaxy has been studied in detail in
the NIR by a number of previous workers most recently Krabbe et al. (2000).
Our CGS4 spectrum adds the detection of the 1-0 Q series to the 
literature as previous spectra did not cover wavelength redward of 2.35$\mu$m.
The majority of the lines are detected as extended with the 1-0 S(1) line
found in 8 pixels (4.9$''$) or 2.3~kpc.

\begin{table}
\caption{Line properties extracted from CGS4 spectra}
\begin{tabular}{llc}
%\noalign{\medskip \hrule \medskip} \hline
Cluster   &  line  &  flux  \\    
          &        & (10$^{-15}$ \ergpcmsqps) \\ \hline
%\hline \endfirsthead
%\hline \multicolumn{5}{continued on next page}
%\endfoot
%\hline\endlastfoot
A11       & Pa$\alpha \lambda 1.876$ &  2.16 $\pm$ 0.15 \\
          & 1-0 S(1)$\lambda 2.122$    & 1.14 $\pm$ 0.16 \\
          & 1-0 S(2)$\lambda 2.034$    & 0.26 $\pm$ 0.09 \\
	  & 1-0 S(3)$\lambda 1.9576$   & 0.49 $\pm$ 0.15 \\
          & 1-0 S(5)$\lambda 1.836$   & 0.53 $\pm$ 0.14 \\
          & 1-0 S(7)$\lambda 1.748$   & 0.71 $\pm$ 0.19 \\
          & [FeII]$\lambda$1.644 & 1.18 $\pm$ 0.16\\ \\
A85       & Pa$\alpha \lambda 1.876$ & 0.4 $\pm$ 0.1 \\
	  & 1-0 S(1)$\lambda 2.122$    & $<0.4$ \\ \\
Zw235     &  no lines detected & \\ \\
A262      & 1-0 S(1)$\lambda 2.122$   & 1.14 $\pm$ 0.28 \\
          & 1-0 S(3)$\lambda 1.9576$  & 1.95 $\pm$ 0.67 \\ \\
RXJ0338+09  & Pa$\alpha \lambda 1.876$ & 2.30 $\pm$ 0.14 \\
	  & 1-0 S(1)$\lambda 2.122$   & 1.21 $\pm$ 0.10 \\
	  & 1-0 S(2)$\lambda 2.034$   & 0.62 $\pm$ 0.06 \\
	  & 1-0 S(3)$\lambda 1.9576$  & 1.81 $\pm$ 0.06 \\ \\
NGC1275   & Pa$\alpha \lambda 1.876$ & 65.7 $\pm$ 0.7 \\
          & Br$\gamma \lambda 2.166$ & 4.4 $\pm$ 0.35 \\
          & Br$\delta \lambda 1.945$ & 4.1 $\pm$ 0.25 \\
	  & 1-0 S(0)$\lambda 2.224$   & 8.44 $\pm$ 0.30 \\
	  & 1-0 S(1)$\lambda 2.122$   & 29.1 $\pm$ 0.36 \\
	  & 1-0 S(2)$\lambda 2.034$   & 7.59 $\pm$ 0.26 \\
	  & 1-0 S(3)$\lambda 1.9576$  & 26.6 $\pm$ 0.34 \\
	  & 1-0 S(4)$\lambda 1.892$   & 7.0 $\pm$ 0.40 \\
	  & 2-1 S(1)$\lambda 2.2477$  & 2.98 $\pm$ 0.30 \\
	  & 2-1 S(3)$\lambda 2.074$  & 2.37 $\pm$ 0.33 \\
	  & 1-0 Q(1)$\lambda 2.407$   & 24.2 $\pm$ 0.53 \\
	  & 1-0 Q(2)$\lambda 2.413$   & 7.86 $\pm$ 0.43 \\
	  & 1-0 Q(3)$\lambda 2.424$   & 23.7 $\pm$ 0.60 \\
	  & 1-0 Q(4)$\lambda 2.438$   & 8.14 $\pm$ 0.46 \\
	  & 1-0 Q(5)$\lambda 2.455$   & 15.3 $\pm$ 0.85 \\ 
          & [SXI]$\lambda$1.92 & 5.2 $\pm$ 0.44 \\ \\
RXJ0352+19 & Pa$\alpha \lambda 1.876$ & 1.58 $\pm$ 0.13 \\ 
	  & 1-0 S(1)$\lambda 2.122$   & 1.29 $\pm$ 0.13 \\
	  & 1-0 S(2)$\lambda 2.034$   & 0.57 $\pm$ 0.097 \\
	  & 1-0 S(3)$\lambda 1.9576$  & 1.85 $\pm$ 0.11 \\
	  & 1-0 S(4)$\lambda 1.892$   & 0.29 $\pm$ 0.15 \\
	  & 1-0 S(5)$\lambda 1.836$  & 0.62 $\pm$ 0.13 \\ \\
RXJ0439+05 & Pa$\alpha \lambda 1.876$ & 0.86 $\pm$ 0.16 \\
	  & 1-0 S(3)$\lambda 1.9576$  & 0.45 $\pm$ 0.16 \\
          & [FeII]$\lambda$1.644 & 0.38 $\pm$ 0.09 \\ \\
RXJ0747-19 & Pa$\alpha \lambda 1.876$ &  2.98 $\pm$ 0.10 \\
	 Br$\gamma 2.166$&  0.25 $\pm$ 0.07 \\
	 & 1-0 S(0)$\lambda 2.224$   &  0.36 $\pm$ 0.10 \\
	 & 1-0 S(1)$\lambda 2.122$   &  1.29 $\pm$  0.09 \\
	 & 1-0 S(2)$\lambda 2.034$   &  0.46 $\pm$ 0.07 \\
	 & 1-0 S(3)$\lambda 1.9576$  &  1.24 $\pm$ 0.085 \\
	 & 1-0 S(4)$\lambda 1.892$   &  0.43 $\pm$ 0.08 \\
	 & 1-0 S(5)$\lambda 1.836$  &  0.92 $\pm$ 0.09\\
	 & 1-0 S(7)$\lambda 1.748$  &  0.50 $\pm$ 0.10 \\
	 & [Si VI]$\lambda1.962$ & 0.50 $\pm$ 0.06 \\ 
         & [FeII]$\lambda$1.644 & 3.0 $\pm$ 0.4 \\ \\
RXJ0821+07            & Pa$\alpha \lambda 1.876$ & 2.60 $\pm$ 0.08 \\
(PA=0 deg)	    Br$\gamma \lambda 2.166$ & 0.27 $\pm$ 0.08 \\ 
		    & 1-0 S(3)$\lambda 1.9576$  & 0.27 $\pm$0.10 \\ \hline
\end{tabular}
\end{table}

\newpage
\begin{table}
\begin{tabular}{llc}
Table 2 continued \\
Cluster   &  line  &    flux  \\    
          &        &   (10$^{-15}$ \ergpcmsqps) \\ \hline
RXJ0821+07            & Pa$\alpha \lambda 1.876$ & 0.80 $\pm$ 0.08 \\
(PA=77 deg)	    & 1-0 S(5)$\lambda 1.836$  & 0.28 $\pm$0.17 \\
                    & [FeII]$\lambda$1.644 & $<$1.2 \\ 
RXJ0821+07-offset   & Pa$\alpha \lambda 1.876$ & 2.37 $\pm$ 0.13 \\ 
(PA=77 deg)         & [FeII]$\lambda$1.644 & $<$0.56 \\ \\
A646      & Pa$\alpha \lambda 1.876$ & 0.52 $\pm$ 0.08 \\
	  & 1-0 S(3)$\lambda 1.9576$  & 0.22 $\pm$ 0.08  \\
	  & 1-0 S(5)$\lambda 1.836$  & 0.27 $\pm$ 0.09 \\ \\
4C+55.16  & Pa$\alpha \lambda 1.876$ & 1.62 $\pm$ 0.16 \\
	  & 1-0 S(3)$\lambda 1.9576$  & 0.85 $\pm$ 0.16 \\
	  & 1-0 S(4)$\lambda 1.892$   & 0.27 $\pm$ 0.13 \\
          & 1-0 S(5)$\lambda 1.836$  & 0.29 $\pm$ 0.08 \\
          & 1-0 S(7)$\lambda 1.748$  & 0.43 $\pm$ 0.08 \\
	  & [FeII]$\lambda$1.644 & 0.86 $\pm$ 0.16 \\ \\
A795      & Pa$\alpha \lambda 1.876$ & 0.52 $\pm$ 0.16 \\ \\
Hydra-A   & Pa$\alpha \lambda 1.876$ & 1.37 $\pm$ 0.14 \\
	  & 1-0 S(1)$\lambda 2.122$   & 0.36 $\pm$ 0.07 \\
	  & 1-0 S(2)$\lambda 2.034$   & 0.30 $\pm$ 0.08 \\	 
	  & 1-0 S(3)$\lambda 1.9576$  & 0.79 $\pm$ 0.12 \\ 
	  & [FeII]$\lambda$1.644 & 1.7 $\pm$ 0.7 \\ \\
Zw3146    & Pa$\alpha \lambda 1.876$ & 2.32 $\pm$ 0.10 \\
          & Pa$\beta \lambda 1.282$  & 1.2 $\pm$ 0.2 \\
	  & 1-0 S(4)$\lambda 1.892$   & 0.35 $\pm$ 0.11 \\
	  & 1-0 S(5)$\lambda 1.836$  & 0.49 $\pm$ 0.09 \\
	  & 1-0 S(7)$\lambda 1.748$  & 0.25 $\pm$ 0.07 \\
	  & [FeII]$\lambda$1.644 & 0.64 $\pm$ 0.08 \\ 
	  & [FeII]$\lambda$1.258 & 1.1 $\pm$ 0.2 \\ \\
A1068     & Pa$\alpha \lambda 1.876$ & 6.57 $\pm$ 0.09 \\
          & Br$\gamma \lambda 2.166$ & 0.66 $\pm$ 0.11 \\
	  & 1-0 S(1)$\lambda 2.122$   & 2.54 $\pm$ 0.09 \\
	  & 1-0 S(2)$\lambda 2.034$   & 0.84 $\pm$ 0.08 \\
	  & 1-0 S(3)$\lambda 1.9576$  & 2.38 $\pm$ 0.10 \\
	  & 1-0 S(4)$\lambda 1.892$   & 0.48 $\pm$ 0.10 \\
	  & 1-0 S(5)$\lambda 1.836$  & 1.38 $\pm$ 0.10 \\
          & 1-0 S(6)$\lambda 1.788$  & 0.37 $\pm$ 0.15 \\
	  & 1-0 S(7)$\lambda 1.748$  & 1.33 $\pm$ 0.09 \\
	  & 2-1 S(3)$\lambda 2.074$  & 0.34 $\pm$ 0.16 \\
	  & [Si VI]$\lambda1.962$ & 0.47 $\pm$ 0.10 \\ 
	  & [FeII]$\lambda$1.644 &  2.5 $\pm$ 0.4 \\ \\
Zw3916   & 1-0 S(7)$\lambda 1.748$& 0.25 $\pm$ 0.12 \\ \\
A1664    & Pa$\alpha \lambda 1.876$ & 2.34 $\pm$ 0.18 \\
	 & 1-0 S(1)$\lambda 2.122$   & 0.79 $\pm$ 0.12 \\
	 & 1-0 S(2)$\lambda 2.034$   & 0.21 $\pm$ 0.07 \\
	 & 1-0 S(3)$\lambda 1.9576$  & 0.60 $\pm$ 0.10 \\
         & 1-0 S(5)$\lambda 1.836$  & 0.42 $\pm$ 0.16 \\ 
         & 1-0 S(7)$\lambda 1.748$  & 0.41 $\pm$ 0.15 \\ \\
          & [FeII]$\lambda$1.644 & 1.06 $\pm$ 0.36  \\ \\
A1795    & Pa$\alpha \lambda 1.876$ & 1.07 $\pm$ 0.13 \\
	 & 1-0 S(1)$\lambda 2.122$   & 0.59 $\pm$ 0.21 \\
	 & 1-0 S(3)$\lambda 1.9576$  & 1.04 $\pm$ 0.09 \\	
	 & 1-0 S(5)$\lambda 1.836$  & 1.19 $\pm$ 0.16 \\ 
          & [FeII]$\lambda$1.644 & 0.5 $\pm$ 0.3  \\ \hline
\end{tabular}
\end{table}

\newpage
\begin{table}
\begin{tabular}{llc}
Table 2 continued \\
Cluster   &  line   &  flux  \\    
          &         & (10$^{-15}$ \ergpcmsqps) \\ \hline
A1835    & Pa$\alpha \lambda 1.876$ & 3.03 $\pm$ 0.09 \\
         & 1-0 S(3)$\lambda 1.9576$  & 0.94 $\pm$ 0.11 \\
         & 1-0 S(5)$\lambda 1.836$  & 0.55 $\pm$ 0.06 \\
         & [FeII]$\lambda$1.644 & 0.34 $\pm$ 0.10 \\ \\
A1885    & no lines detected & \\ \\
Zw7160 & [FeII]$\lambda$1.644 & 0.18 $\pm$ 0.07 \\ \\
A2029  & lineless control  & \\ \\
A2052     & Pa$\alpha \lambda 1.876$ & 0.75 $\pm$ 0.24 \\ 
          & [FeII]$\lambda$1.644 & $<$0.55 \\ \\
A2199     & Pa$\alpha \lambda 1.876$       & 0.89 $\pm$ 0.35 \\ \\
A2204     & Pa$\alpha \lambda 1.876$ & 2.17 $\pm$ 0.10 \\
          & Pa$\beta \lambda 1.282$  & 0.6 $\pm$ 0.4 \\
          & 1-0 S(1)$\lambda 2.122$   & 1.18 $\pm$ 0.12 \\
	  & 1-0 S(2)$\lambda 2.034$   & 0.48 $\pm$ 0.09 \\
          & 1-0 S(3)$\lambda 1.9576$  & 1.21 $\pm$ 0.08 \\
          & 1-0 S(4)$\lambda 1.892$   & 0.36 $\pm$ 0.08 \\
          & 1-0 S(5)$\lambda 1.836$  & 0.78 $\pm$ 0.08 \\
          & 1-0 S(7)$\lambda 1.748$  & 0.43 $\pm$ 0.10 \\
          & 2-1 S(3)$\lambda 2.074$  & 0.17 $\pm$ 0.07 \\
          & [FeII]$\lambda$1.644 & 2.0 $\pm$ 0.4  \\ 
          & [FeII]$\lambda$1.258 & 2.1 $\pm$ 0.3  \\ \\
Her-A     & Pa$\alpha \lambda 1.876$ & 0.37 $\pm$ 0.08 \\
          & 1-0 S(1)$\lambda 2.122$   & 0.28 $\pm$ 0.10 \\
          & 1-0 S(3)$\lambda 1.9576$  & 0.35 $\pm$ 0.08 \\
          & 1-0 S(5)$\lambda 1.836$  & 0.21 $\pm$ 0.06 \\ \\
Zw8193    & Pa$\alpha \lambda 1.876$  & 1.42 $\pm$ 0.08 \\
(Component A)          & 1-0 S(3)$\lambda 1.9576$   & 0.79 $\pm$ 0.10 \\
	  & [FeII]$\lambda$1.644 & 1.12 $\pm$ 0.13 \\ 
Zw8193    & Pa$\alpha \lambda 1.876$ & 1.07 $\pm$ 0.09 \\
(Component B)		     & [FeII]$\lambda$1.644 & 0.69 $\pm$ 0.19 \\ \\
Zw8197    & no lines detected    & \\ \\
Zw8276    & Pa$\alpha \lambda 1.876$  & 0.43 $\pm$ 0.09 \\
          & 1-0 S(1)$\lambda 2.122$ & 0.82 $\pm$ 0.09 \\
	  & 1-0 S(3)$\lambda 1.9576$& 1.02 $\pm$ 0.09 \\
	  & 1-0 S(5)$\lambda 1.836$& 0.47 $\pm$ 0.1 \\ \\
A2390     & Pa$\alpha \lambda 1.876$  & 0.43 $\pm$ 0.09 \\
	  & 1-0 S(3)$\lambda 1.9576$& 0.45 $\pm$ 0.13 \\
	  & [FeII]$\lambda1.644$ & 0.38 $\pm$ 0.16\\ \\
A2597     & Pa$\alpha \lambda 1.876$ & 2.43 $\pm$ 0.20 \\
	  & 1-0 S(0)$\lambda 2.224$ & 1.08	$\pm$ 0.18 \\
	  & 1-0 S(1)$\lambda 2.122$ & 3.38 $\pm$ 0.10 \\
	  & 1-0 S(2)$\lambda 2.034$ & 1.20 $\pm$ 0.19 \\
	  & 1-0 S(3)$\lambda 1.9576$& 3.34 $\pm$ 0.15 \\
	  & 1-0 S(4)$\lambda 1.892$ & 0.82 $\pm$ 0.17 \\
	  & 1-0 S(5)$\lambda 1.836$& 2.00 $\pm$ 0.14 \\
	  & 1-0 S(7)$\lambda 1.748$& 1.05 $\pm$ 0.21 \\ 
	  & 2-1 S(1)$\lambda 2.2477$& 0.77 $\pm$ 0.30 \\ \hline
%\noalign{\smallskip \hrule}
\end{tabular}
\end{table}
%\vfill\eject
%\end{longtable}

% missing objects + lines:
%
% A2199 and Her-A and odd lines for 0821
%
% FeII for    1795, 2052 and 2204
%
%

\begin{figure*}
\vspace*{-1cm}
\centerline{\psfig{file=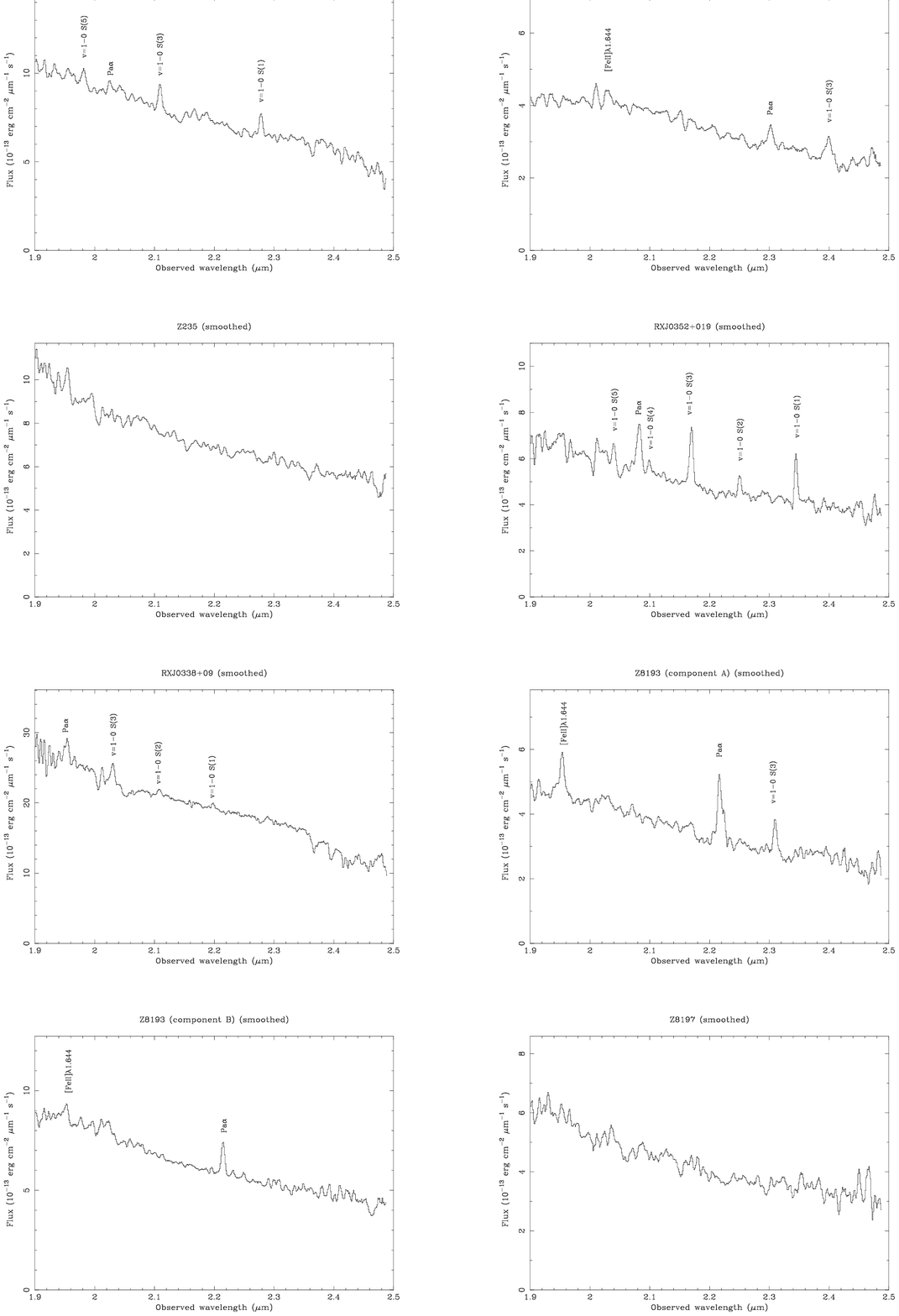,angle=0,width=18cm}}
\vspace*{-1cm}
\caption{K-band spectra for the sample.}
\end{figure*}

\begin{figure*}
\addtocounter{figure}{-1}
\vspace*{-1cm}
\centerline{\psfig{file=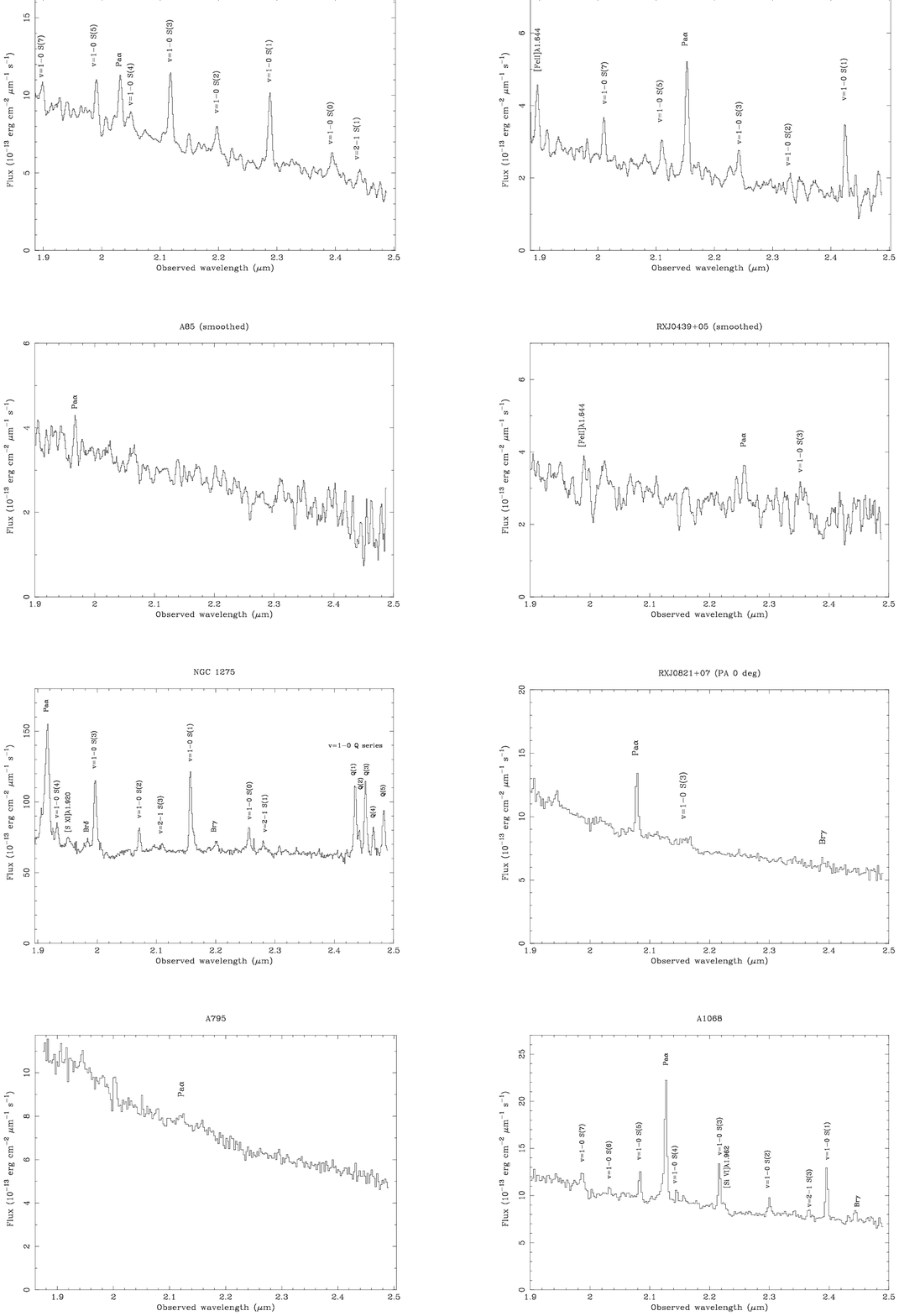,angle=0,width=18cm}}
\vspace*{-1cm}
\caption{K-band spectra for the sample continued.}
\end{figure*}

\begin{figure*}
\addtocounter{figure}{-1}
\vspace*{-1cm}
\centerline{\psfig{file=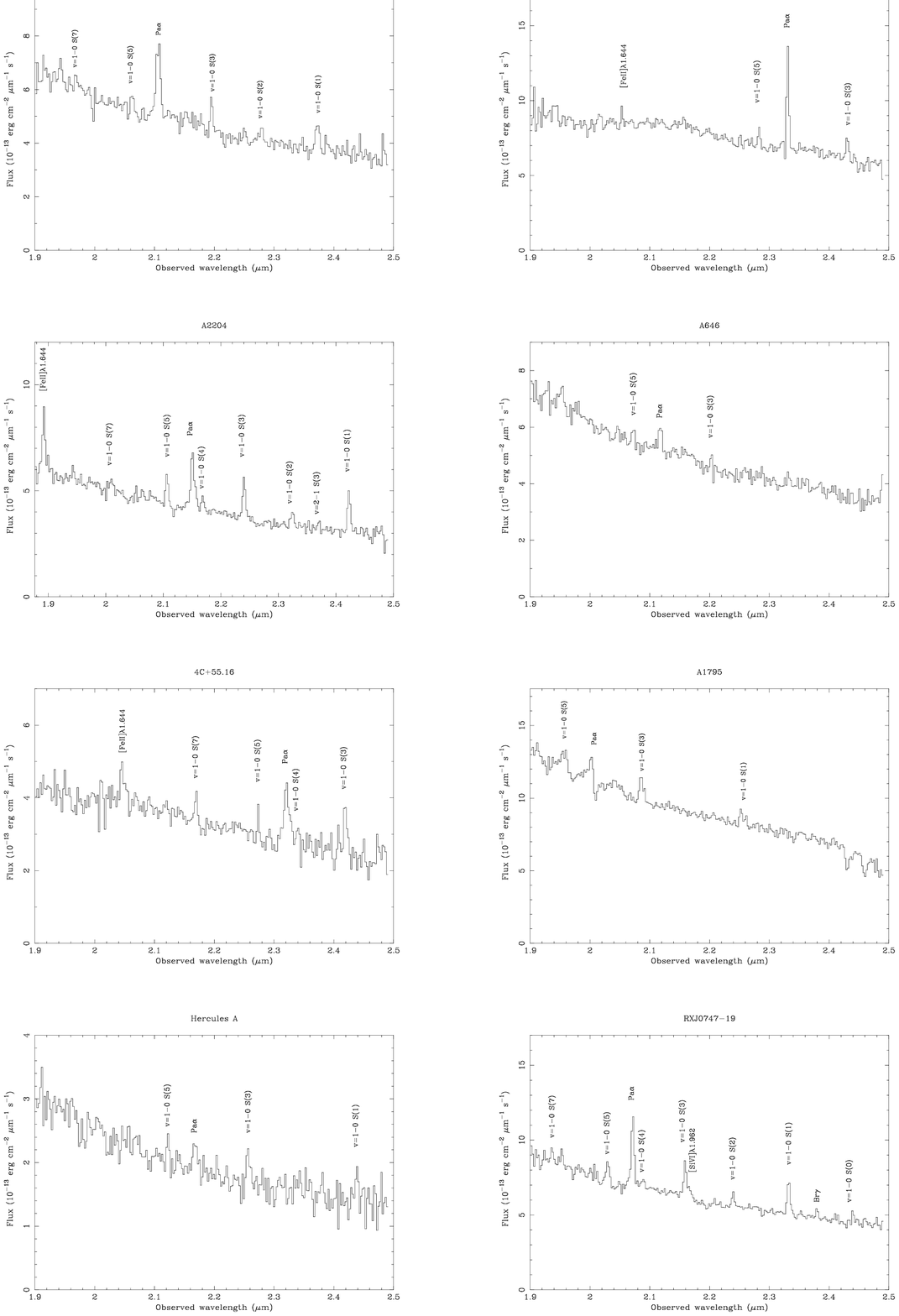,angle=0,width=18cm}}
\vspace*{-1cm}
\caption{K-band spectra for the sample continued.}
\end{figure*}

\begin{figure*}
\addtocounter{figure}{-1}
\vspace*{-1cm}
\centerline{\psfig{file=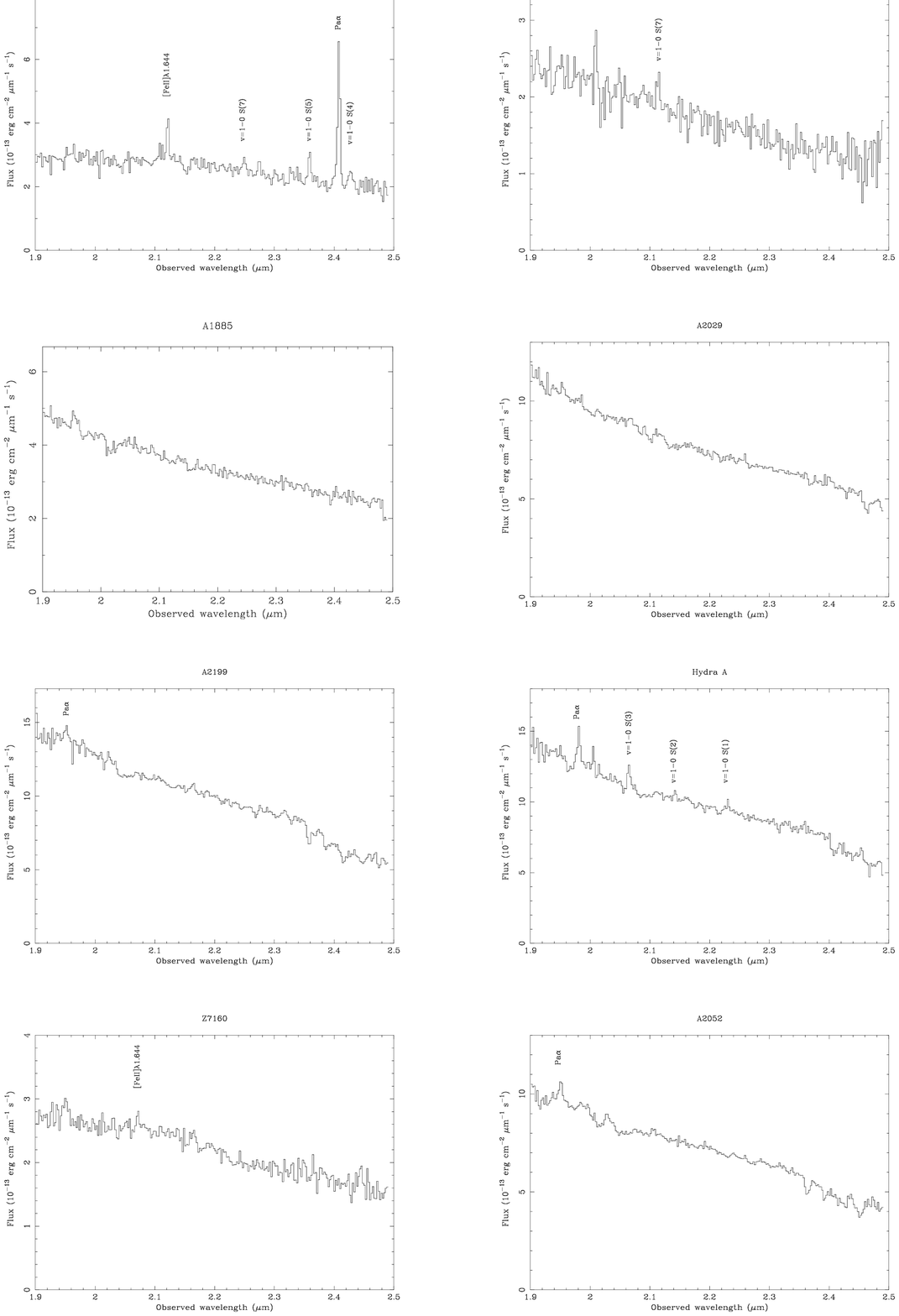,angle=0,width=18cm}}
\vspace*{-1cm}
\caption{K-band spectra for the sample continued.}
\end{figure*}

\begin{figure*}
\vspace*{-3cm}
\centerline{\psfig{file=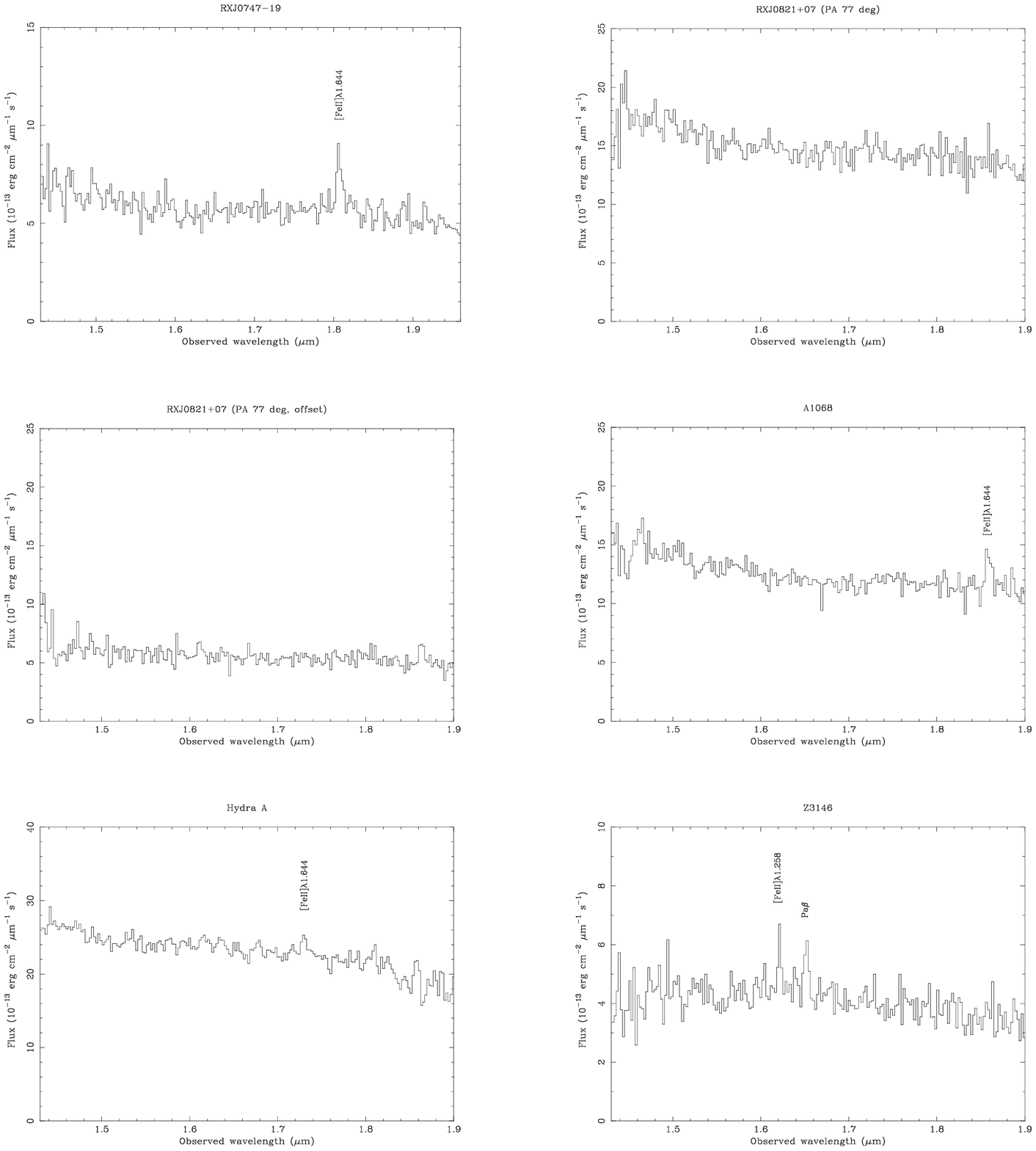,angle=0,width=18cm}}
\vspace*{-3cm}
\caption{H-band spectra for the sample.}
\end{figure*}

\begin{figure*}
\addtocounter{figure}{-1}
\vspace*{-6cm}
\centerline{\psfig{file=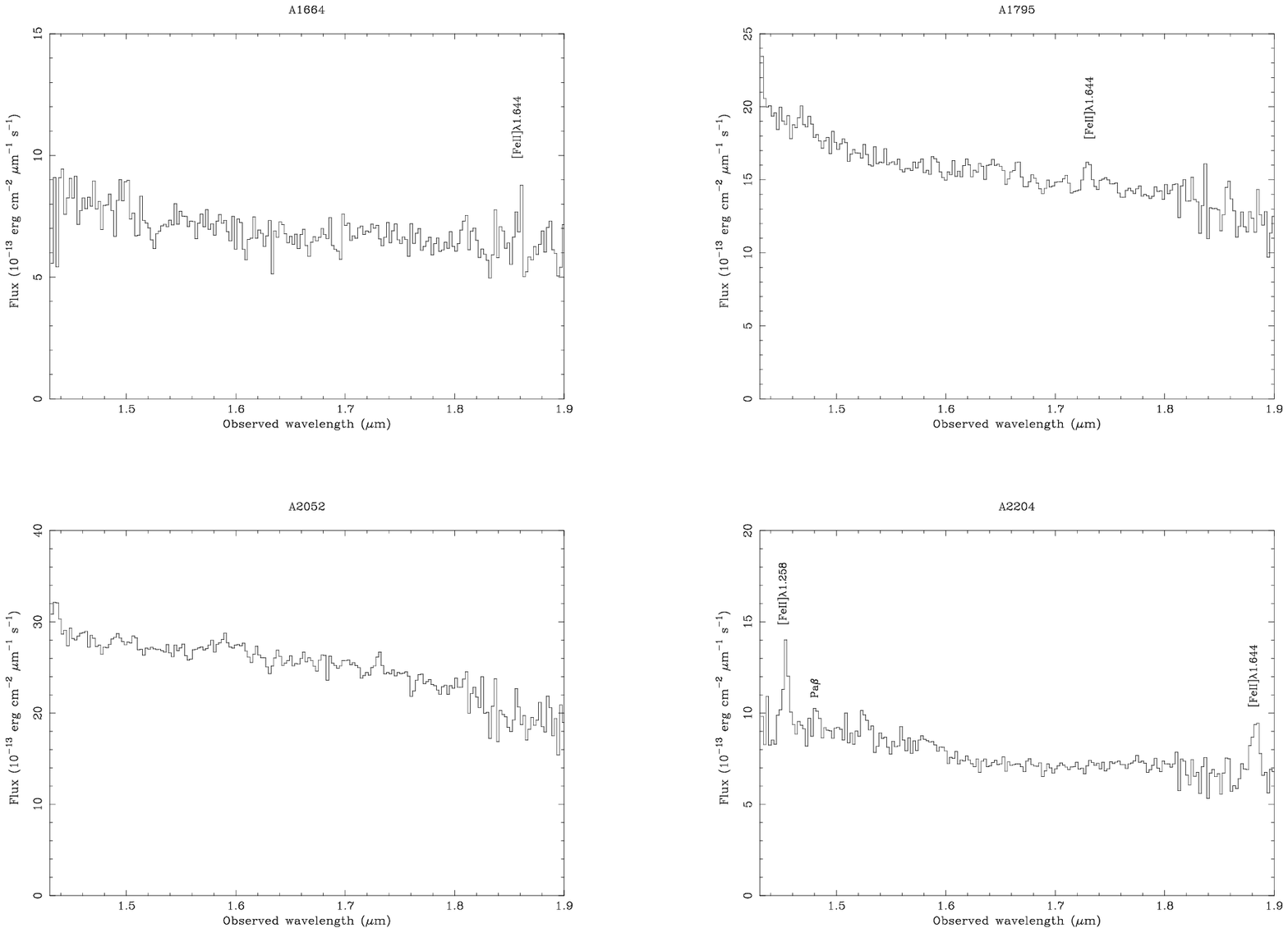,angle=0,width=18cm}}
\vspace*{-6cm}
\caption{H-band spectra for the sample continued.}
\end{figure*}

\section{DISCUSSION}

The observations presented here are the largest near-IR spectroscopic study 
of central cluster galaxies in cooling flows by a factor of four.
Seven of the objects presented have previously published IR spectra.
Hydra-A, RXJ0338+09, RXJ0747-19, NGC1275, A1795, A2029  and A2597 are included in Falcke et al. (1998),
Jaffe \& Bremer (1997), Krabbe et al. (2000) and Jaffe, Bremer \& van der Werf (2001). 
The line fluxes are all within $<$30\% of those presented elsewhere
when the different slit widths are taken into account (e.g. Jaffe et al.\ 2001 use a
slit width of 2$''$ to 1.2$''$ used in the paper).
In addition, this study includes, for the first time, coverage of the [FeII] line.

In this section we discuss the limits this dataset can place on
reddening, trends in the [FeII] line, detection of high ionization lines
and column density constraints.

\subsection{Reddening}

One of the most important results from the optical spectroscopy of central
cluster galaxies is the significant reddening of the lines (Allen et al.\ 1995;
Crawford et al.\ 1999). The values of E(B-V) derived imply column densities of 
1--5$\times 10^{21}$ cm$^{-2}$. 

There are several combinations of optical and NIR lines that can be used
to gauge the reddening. The most straightforward is H$\alpha$/Pa$\alpha$. 
Figure 3 shows the ratio of  H$\alpha$/Pa$\alpha$ with redshift for our
joint detections with Crawford et al. (1999). The majority of the points
agree with the ratio expected from Case B recombination (Osterbrock 1989)
but several outliers are obvious. The majority of the objects have
ratios consistent with E(B-V)$<0.1$ and are consistent with the values
derived from H$\beta$/H$\alpha$ ratios by Crawford et al. (1999). The objects
with the three lowest ratios (A1068, A1835 and RXJ0821+07) have high
E(B-V) values in  Crawford et al. (1999) (0.39$^{+0.07}_{-0.08}$, 0.40$^{+0.06}_{-0.06}$ 
and 1.16$^{+0.33}_{-0.58}$ respectively)
which are consistent with those implied from this analysis
(0.54$\pm$0.02, 0.54$\pm$0.03 and 0.71$\pm$0.04). However, there
are several objects with similar values of E(B-V) in  Crawford et al. (1999)
where our estimate is substantially lower (e.g. A1664, 0.46$^{+0.06}_{-0.07}$  vs. 0.12$\pm$0.05).
This variation may be due in part to
the extent of the emission lines and the fact that the optical
and CGS4 spectra, while using the same slit width (1.2$''$), 
were not performed at the same position angle. Therefore
there are several points (e.g. Zw8276) where the optical line is brighter
by a factor of two and others (e.g. RXJ0821+07) where the line is fainter
by a factor of three.

The only other reddening diagnostic line combination available within our
CGS4 data are the [FeII] lines at 1.258 and 1.644$\mu$m. We have data
for two objects, Zw3146 and A2204, and the equivalent E(B-V) values,
given the expected ratio of 1.258$\mu$m to 1.644$\mu$m of 1.3, are
$<$0.1 and 0.5$\pm$0.3 (as opposed to 0.2 and 0.0 in Crawford et al. 1999).

Given the restrictions on the current datasets it is not possible to make
any quantitative statements about reddening from our data. Future 
narrow-band imaging in the optical and NIR will allow direct comparison of
the H$\beta$, H$\alpha$ and Pa$\alpha$ lines and enable us to derive 
more reliable reddening maps.

\begin{figure*}
\centerline{\psfig{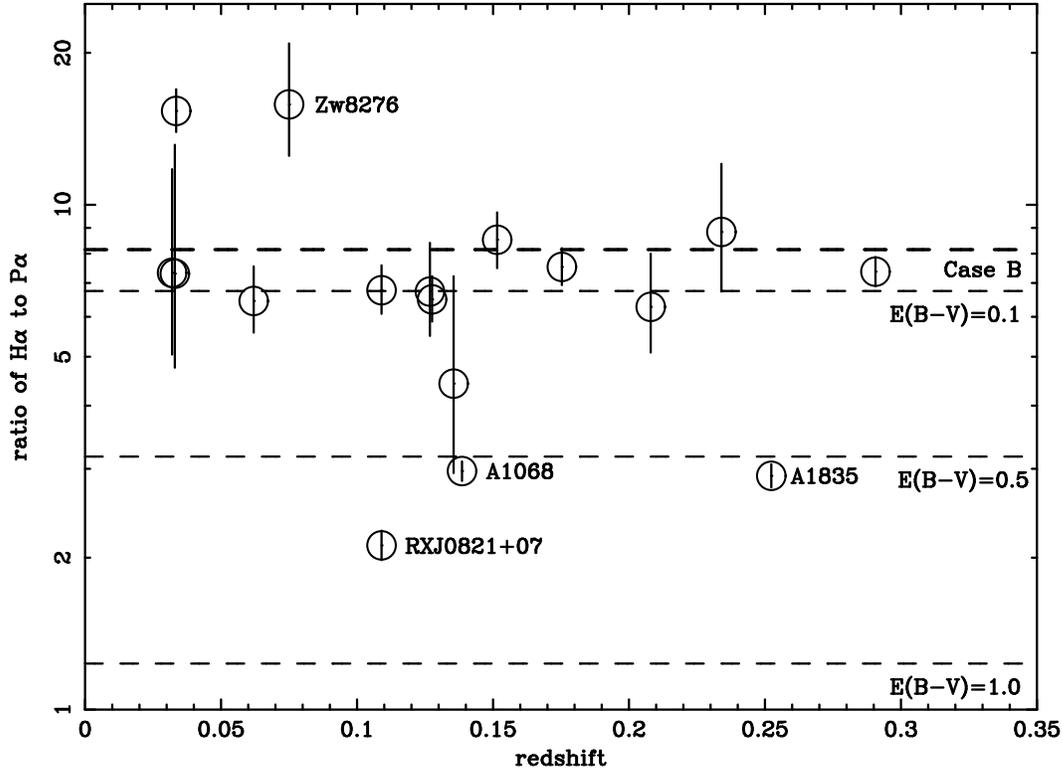}}
\caption{The ratio of the H$\alpha$ to Pa$\alpha$ lines plotted against redshift.
The Case B prediction of 8.153 for T$=$10000K from Osterbrock (1989) is plotted 
for values of reddening of E(B-V)=0.0,0.1,0.5 and 1.0.}
\end{figure*}

\subsection{[FeII] emission}

For the majority of our sample we have obtained data for [FeII]1.644$\mu$m
through the redshifting of the line into the K-band or an H-band spectrum.
Time limitations during our first run prevented all low redshift objects
being observed in the H-band but we have [FeII] significant detections for 12 of a total of 18
objects that we have covered at the equivalent of rest 1.644$\mu$m (and an additional
two marginal detections, A1795 and A2390).
Most of the non-detections for [FeII] are for clusters without 
any molecular line detection in the K-band.

From the [FeII] data we have it appears that the strength of this feature
correlates with the other near-infrared and optical lines but most significantly
with [OI] 6300\AA (Mouri et al. 1990). 
There are several objects where the [FeII] is particularly strong (e.g. A2204
and A1068) and these objects show the strongest [OI]. 
Therefore, these two lines are probably closely related in these systems.
Figure 4 shows a comparison of the fluxes of [OI] and [FeII] lines for this sample.

There is a substantial literature on the potential origin of [FeII]
emission in Seyfert galaxies and starbursts (Kawara, Nishida \& Taniguchi 1988,
Mouri et al. 1990, Simpson et al. 1996, Mouri, Kawara, \& Taniguchi  2000) but no firm
conclusions have yet been drawn on the exact excitation mechanism 
in these systems. The [FeII] and [OI] are of particular diagnostic importance
as they have ionization potentials of 16.2 and 13.6eV respectively so trace regions
where hydrogen is partially ionized as it has a very similar ionization potential.
Shocks or X-ray heating are favoured in cases where [FeII] is 
strong compared to Br$\gamma$ and both these mechanisms
may play a role in the cooling flow systems presented here (see Wilman et al. 2002).

 \begin{figure*}
\centerline{\psfig{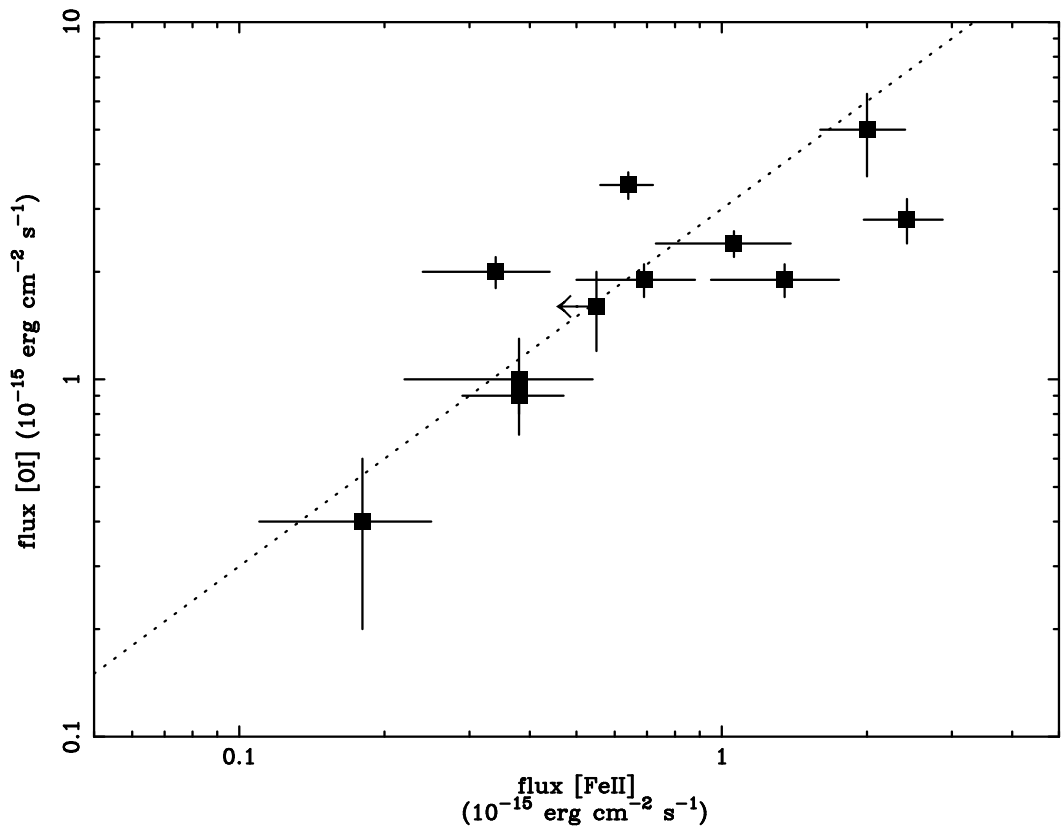}}
\caption{The flux of the [OI] 6300\AA \ line plotted against that of [FeII] 1.644$\mu$m.
The dotted line marks the best least-squares linear fit (with an index fixed at
1.0) to the data points.}

\end{figure*}

\subsection{[SiVI] and other high ionization lines}

One of the most striking results from this study is the presence of
[SiVI] and several other  high ionization lines (e.g. [SIX], 2-1 S(3)
and HeI in the wing of Pa$\alpha$)
in a few objects (Cyg-A; Wilman et al.\ (2000), A1068 and PKS0745-191).
These objects are the ones with the strongest [OIII] lines
(e.g. all central cluster galaxies known with [OIII] brighter
than 10$^{-14}$ erg cm$^{-2}$ s$^{-1}$)
and most (but not all) have strong radio sources.
The ionization potential of [SiVI] (167~eV) implies the
presence of a strong non-thermal source and is consistent
with the optical and radio observations. 

One prediction from these observations is that high 
resolution X-ray imaging may detect the unresolved X-ray
emission from the nucleus (as found in Hydra-A - McNamara et al.\ 2000).
All the objects in this study with detected [SiVI] have
{\it Chandra} observations and the brightest (NGC~1275) has a 
well-established nuclear detection. 

Our data are at too low a spectral resolution to determine if the
high ionization lines have a significantly different line
width to the Paschen or molecular hydrogen lines. In addition,
we can determine no spatial differences in these lines. 
Clearly higher spectral and spatial resolution data will help
determine the relative importance of nuclear excitation 
in different objects. The orientation of any radio jets
and dust lanes in {\it HST} images are additional factors in
determining what is illuminated and how much of it we observe.

As noted above, the presence of higher ionization lines is
not necessarily related to the radio power of the central 
galaxy. A1068 is a relatively weak radio source but has
strong [OIII] and significant Br-$\gamma$ and [Si VI].
Therefore, the active nuclei in central cluster galaxies
are not exclusively radio-loud.

\subsection{Spatial and velocity structure}

A number of galaxies in this study show extended line 
emission (as discussed in section 3). The four systems
with the most significant extents are presented in Figure 5,
which shows the line strength of 1-0 S(1) and Pa$\alpha$
with position on the slit.
Three of these four show stronger Pa$\alpha$ in the centre
compared to  1-0 S(1), with only A2597 showing a constant
line ratio. This spatial variation implies that the 
excitation of hydrogen is strongest near the nucleus of
the galaxy, probably due to the action of an active nucleus
as discussed above. Variations in 1-0 S(1) to H$\alpha$+[NII]
of a similar magnitude are observed by Donahue et al. (2000)
for A2597 and RXJ0747-19. Although the effects of dust could
significantly affect the Donahue et al. (2000) ratios, it
is encouraging to note that spatial variations can be extracted
from {\sl HST} imaging of such systems and hence the relative
importance of different excitation mechanisms with radius
from the centre of the galaxy.

Two of the galaxies in this study also show significant
velocity structure (A2204 and Zw8193). Figure 6 shows
the reduced 2-D frames of these objects. The
velocity shear is present in both the ionized and molecular
gas lines in A2204  but is most pronounced in the
1-0 S(3) line. In Zw8193 the molecular lines are weak
compared to Pa$\alpha$ so a direct comparison of the
two phases is not possible.

The amplitudes of these velocity shifts ($>$300~km~s$^{-1}$)
are at the limit of the resolution provided by CGS4 and higher
resolution integral field spectroscopy has the potential to
reveal more about the properties of these systems.
Determining the velocity structure of the ionized and
molecular gas components can set stringent limits
on the importance of galaxy-galaxy or ICM/radio-jet interactions
in these dense cooling flow core regions.

\begin{figure*}
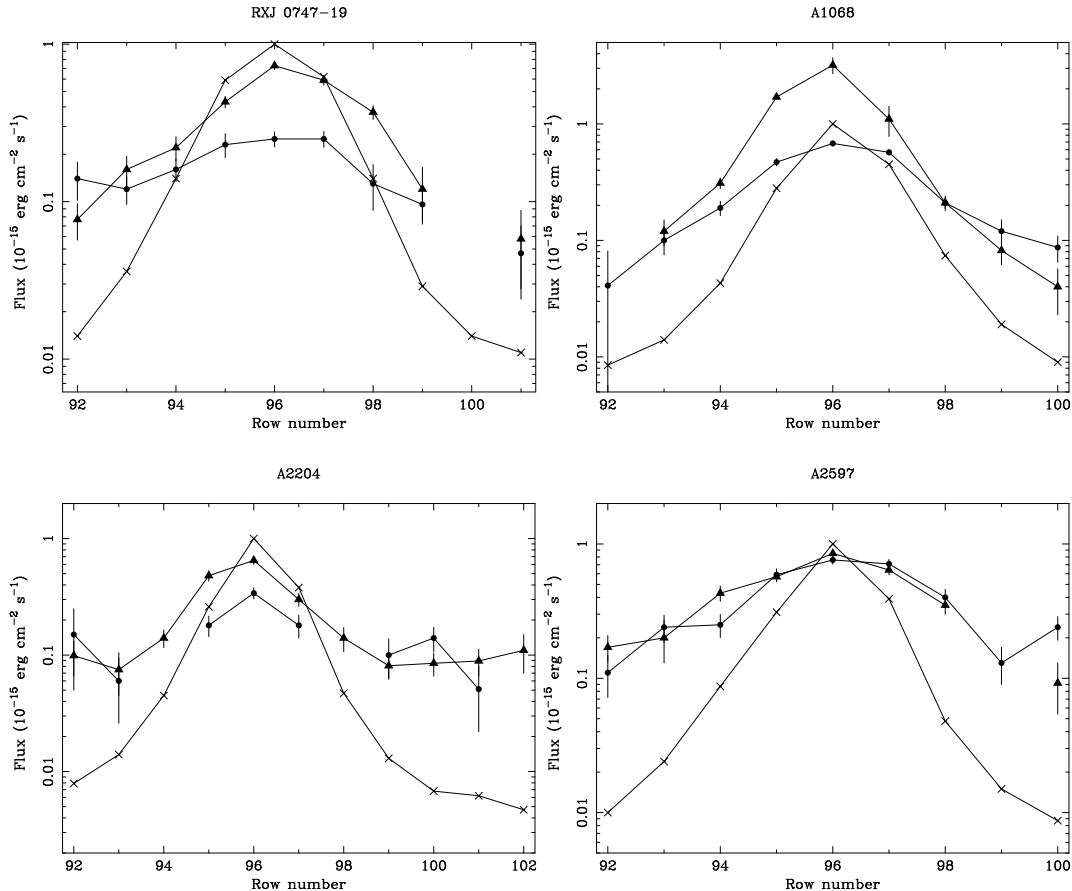

\centerline{\psfig{file=fig5a.ps,angle=270,width=7cm}
\psfig{file=fig5b.ps,angle=270,width=7cm}}
\vspace*{0.5cm}
\centerline{\psfig{file=fig5c.ps,angle=270,width=7cm}
\psfig{file=fig5d.ps,angle=270,width=7cm}}
\caption{Variation of line flux for 1-0 S(1) (filled circles) and 
Pa$\alpha$ (filled triangles) with
row number (each row corresponds to 0.6$''$) for the four
most significantly extended systems. Instrumental point
spread functions, as derived from contemporaneous standard 
star observations, are marked with crosses. Where no point
is plotted there is no significant detection.}
\end{figure*}

\begin{figure*}
\vspace*{-0.5cm}
\centerline{\psfig{file=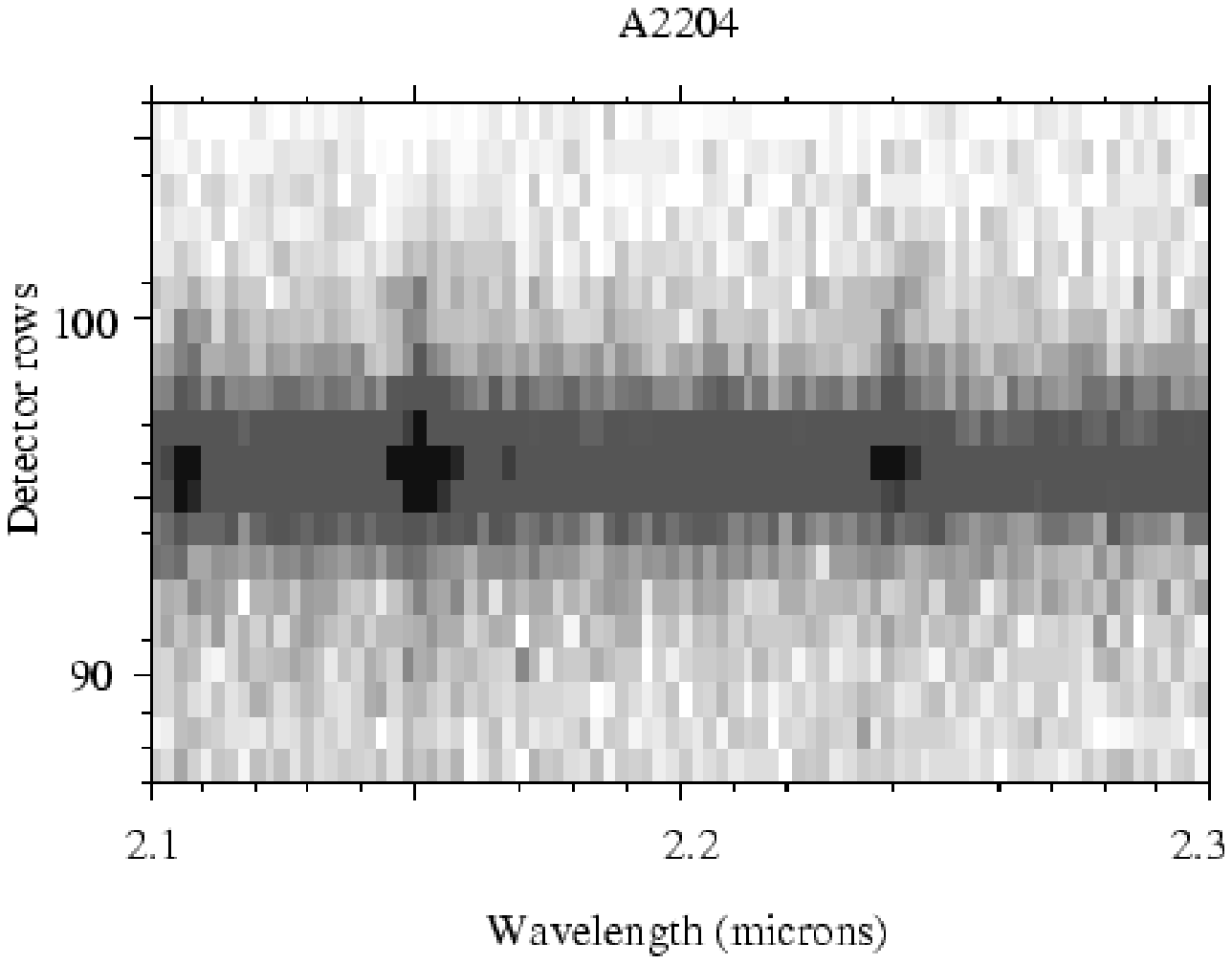,angle=270,width=14cm}}
\vspace*{-1cm}
\centerline{\hspace*{1.8cm} \psfig{file=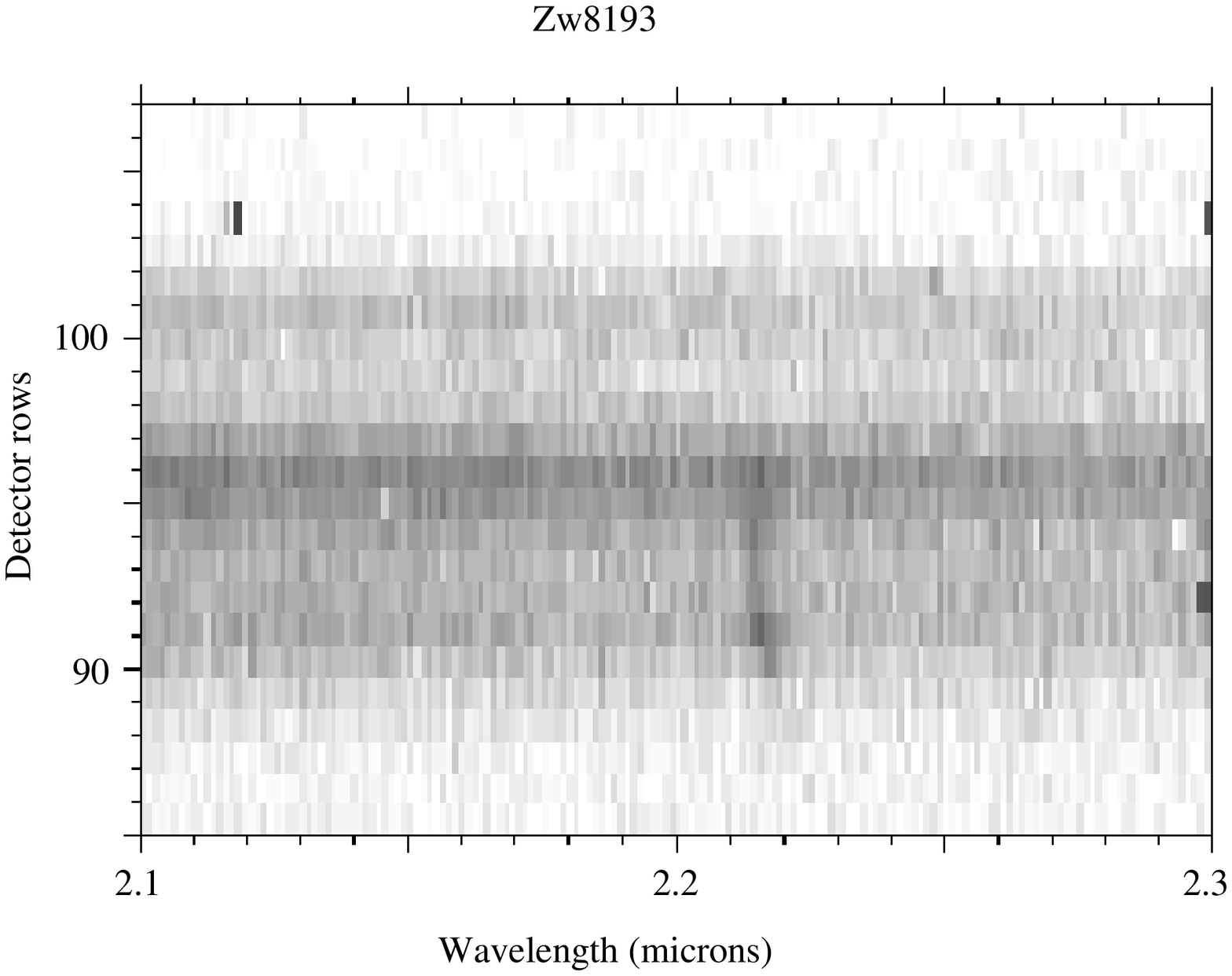,angle=270,width=20cm}}
\vspace*{-1.5cm}
\caption{A portion of the reduced, two-dimensional spectral image
for A2204 and Zw8193 showing the velocity structure in the
lines of Pa$\alpha$ (2.15$\mu$m) and 1-0 S(3) (2.25$\mu$m) for A2204
and Pa$\alpha$ (2.22$\mu$m) for Zw8193. }
\end{figure*}

\subsection{Correlation with CO-deduced molecular gas mass}
 
The recent detection of CO in many of the targets in this
survey by Edge (2001) allows us to compare the amount
of `hot' ($>$1000~K) to `cool' ($\sim$30~K) molecular
hydrogen. Figure 7 shows the luminosity of the 1-0 S(1)
line (or its equivalent) against molecular hydrogen mass
using data from this paper, Cygnus-A from Wilman et al. (2000)
and A478 from Jaffe et al. (2001). The luminosity is 
determined for the slit flux only so the overall normalisation
of the correlation may vary for the more extended emission. 
There is an apparent correlation similar to that found in Edge (2001)
for the luminosity of H$\alpha$. To confirm the validity of these
luminosity-luminosity correlations we calculated the 
Spearmans rho statistic (as recommended by Feigelson \& Berg 1983)
using the ASURV package
for both the 1-0 S(1) and H$\alpha$ luminosities correlated
against the molecular hydrogen mass estimated from CO. The
correlation is significant for CO vs. 1-0 S(1) ($\rho$=0.59, implying
a probability of correlation of $>99.5$ per cent).
The result for CO vs. H$\alpha$ is even more significant ($\rho$=0.63, so 
P$>99.7$ per cent). This
result improves if Cygnus-A is removed from the analysis
($\rho$=0.72). As pointed out by  Feigelson \& Berg (1983),
flux-flux correlations are not as reliable at determining whether
an underlying correlation is present particularly for samples
where most detections are close to the detection limit (as is the case in
this study). We did perform flux-flux correlations and get
significant correlations for both optical and infrared line correlations
but only if Perseus is included. The correlations presented in Figure~7
and Figure~9 of Edge (2001) are significant and are not an
artifact.

It is interesting to note the relatively low fluxes of 1-0 S(1) 
and/or S(3) in the most radio luminous objects. The comparison
with  H$\alpha$ shows the objects with powerful radio sources which 
have firm CO upper limits (e.g. Cygnus-A and RXJ0747-19) lie much 
further from the correlation between CO and optical
line luminosity than for the luminosity of 1-0 S(1) as a result
of the much higher ratios of  H$\alpha$/1-0 S(1) found for these
objects. Figure~7 demonstrates that the interelation between CO and 
the 1-0 S(1) line may be closer than that with ionized hydrogen but 
as yet the observational datasets
are not large enough or accurate enough to confirm this.

As with the high ionization lines, the spectral resolution of
our CGS4 data is not sufficient to obtain line widths for the
H$_2$ lines so we cannot draw a direct comparison between these
two phases in velocity. The CO lines have widths between 
150 and 500~km~s$^{-1}$
so  higher resolution near-infrared spectroscopy would allow
the line profiles to be compared to establish if cool and warm components
of molecular hydrogen are bound in clouds with the same velocity structure.

Irrespective of the excitation mechanism (which are dealt
with in more detail in Wilman et al. 2002),
the results clearly demonstrate that 
in the central cluster galaxies within strong
cooling flows we are observing clouds of dense molecular gas being
warmed  to produce ionized, 
atomic hydrogen (H$\alpha$, Pa$\alpha$),
warm molecular hydrogen in the near-IR and cool molecular hydrogen via CO.
The exact ratios of these three phases will undoubtedly vary with
viewing angle, star-formation rate and strength of any nuclear source
but our observations show that any deviations are not larger than
an order of magnitude. Therefore the scaling apparent in Figure~7 holds
in the majority of cases with the most deviant points offering
the possibility to constrain the excitation mechanisms at work.
Our relative sensitivities to these three phases mean that it is the
H$\alpha$ emission that is most easily detected and the CO the hardest. 
There is also a possibility that a fourth phase, neutral, atomic hydrogen
visible through the 21cm line, should also be present as photo-dissociation
region models predict a `skin' of atomic hydrogen (Kaufman et al.\ 1999). 
This phase will be the
most challenging to detect but would strongly support the view that
all the `peculiar' features of cooling flow central cluster galaxies
stem from the presence of cold molecular clouds. 

\begin{figure*}
\centerline{\psfig{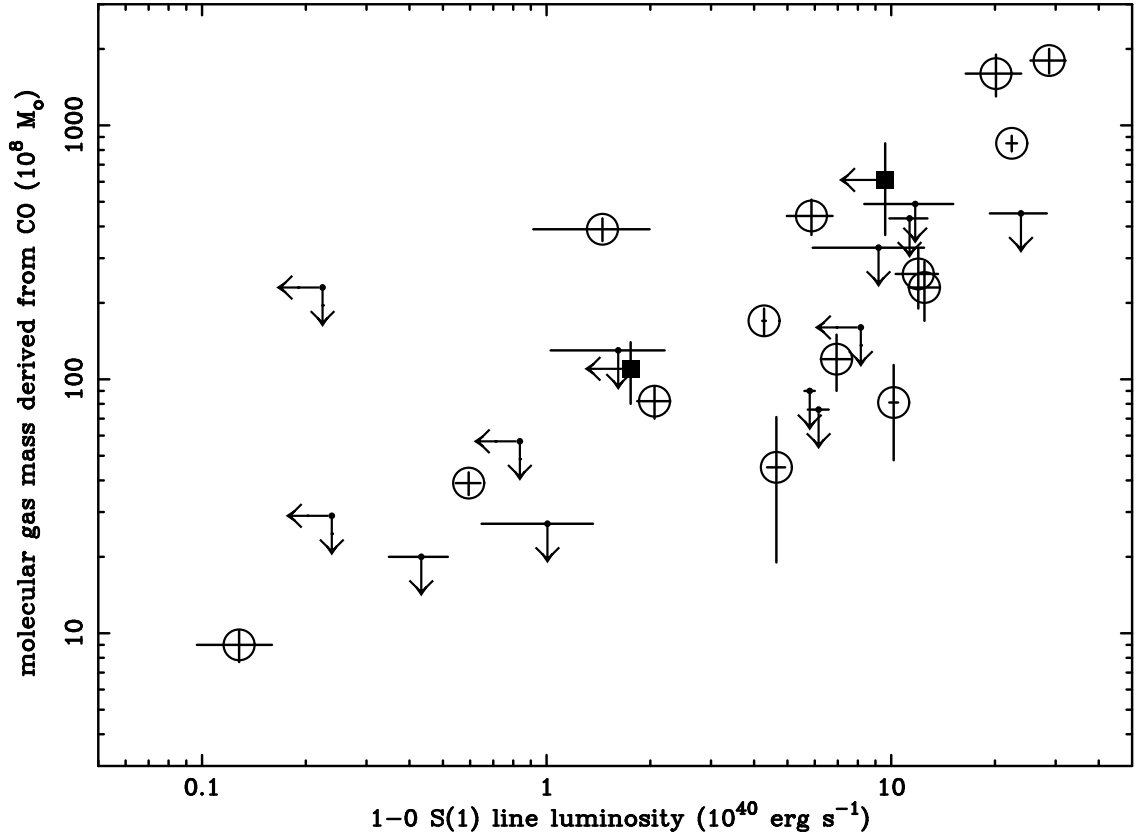}}
\caption{The mass of molecular gas from Edge (2001) plotted against the 
luminosity of the H$_2$ 1-0 S(1)
line (or its equivalent for higher redshift objects). The filled squares mark Zw7160 and
Zw8197 the only CO detections by Edge (2001) without corresponding 1-0 S(1) detections.
The dotted line marks a linear relationship between the two properties and is not a fit
to the data.
}

\end{figure*}

\begin{table*}
\begin{tabular}{lccc}
cluster & H$\alpha$ luminosity    & 1-0 S(1) luminosity      &  molecular gas mass \\
        & from C99 (erg~s$^{-1}$) & (erg~s$^{-1}$) &  from Edge (2001) \\
        &          &                & \\
A11        & (1$\times 10^{42}$)   & 1.2$\pm0.2 \times 10^{41}$  & 2.6$\pm0.7\times 10^{10}$   \\
A85        & --                    & $<5.5  \times 10^{39}$      & --  \\
Zw235      & 4.1$\times 10^{40}$   & $<9.2  \times 10^{39}$      & --  \\
A262       & 5.5$\times 10^{39}$   & 1.3$\pm0.3 \times 10^{39}$  & 9.0$\pm1.3\times 10^{8}$  \\
RXJ0338+09 & 1.8$\times 10^{41}$   & 5.9$\pm0.5 \times 10^{39}$  & 3.9$\pm0.4\times 10^{9}$  \\
NGC1275    & (5$\times 10^{42}$)   & 4.2$\pm0.1 \times 10^{40}$  & ($1.7\pm0.2 \times 10^{10}$)  \\
RXJ0352+19 & 5.9$\times 10^{41}$   & 6.9$\pm0.7 \times 10^{40}$  & 1.2$\pm0.3\times 10^{10}$  \\
A478       & 1.1$\times 10^{41}$   & 4.6$\pm0.3 \times 10^{40}*$  & 4.5$\pm2.6\times 10^{9}$  \\
RXJ0439+05 & 1.1$\times 10^{41}$   & 9.2$\pm3.3 \times 10^{40}$  & $<3.3 \times 10^{10}$  \\
RXJ0747-19 & (1.4$\times 10^{42}$) & 6.2$\pm0.4 \times 10^{40}$  & $<7.6 \times 10^{9}$  \\
RXJ0821+07 & 3.0$\times 10^{41}$   & 1.5$\pm0.5 \times 10^{40}$  & 3.9$\pm0.4 \times 10^{10}$  \\
A646       & 2.6$\times 10^{41}$   & 1.6$\pm0.6 \times 10^{40}$  & $<1.3 \times 10^{10}$  \\
4C+55.16   & (1$\times 10^{42}$)   & 2.3$\pm0.4 \times 10^{41}$  & $<4.5 \times 10^{10}$  \\
A795       & 1.9$\times 10^{41}$   & $<2.5 \times 10^{40}$       & --  \\
Hydra-A    & (1.6$\times 10^{41}$) & 4.3$\pm0.8 \times 10^{39}$  & $<2.0 \times 10^{9}$  \\
Zw3146     & 7.1$\times 10^{42}$   & 2.0$\pm0.4 \times 10^{41}$  & 1.6$\pm0.3 \times 10^{11}$  \\
A1068      & 1.7$\times 10^{42}$   & 2.2$\pm0.1 \times 10^{41}$  & 8.5$\pm0.6 \times 10^{10}$  \\
Zw3916     & 1.5$\times 10^{41}$   & $<8.2 \times 10^{40}$       & $<1.6 \times 10^{10}$  \\
A1664      & 1.1$\times 10^{42}$   & 5.9$\pm0.9 \times 10^{40}$  & 4.4$\pm0.7 \times 10^{10}$  \\
A1795      & 1.2$\times 10^{41}$   & 1.0$\pm0.4 \times 10^{40}$  & $<2.7 \times 10^{9}$  \\
A1835      & 2.7$\times 10^{42}$   & 2.9$\pm0.3 \times 10^{41}$  & $1.8\pm0.2 \times 10^{11}$  \\
A1885      & 5.4$\times 10^{40}$   & $<1.1 \times 10^{40}$       & --  \\
Zw7160     & 5.0$\times 10^{41}$   & $<9.6 \times 10^{40}$       & $6.1\pm2.4 \times 10^{10}$  \\
A2029      & $<1.2\times 10^{40}$  &  $<8.4 \times 10^{39}$       & ($<5.7\times 10^{9}$)  \\
A2052      & 3.0$\times 10^{40}$   & $<2.2 \times 10^{39}$       & ($<2.3\times 10^{10}$) \\
A2199      & 2.8$\times 10^{40}$   & $<2.4 \times 10^{39}$       & ($<2.9\times 10^{9}$)  \\
A2204      & 1.9$\times 10^{42}$   & 1.3$\pm0.1 \times 10^{41}$  & $2.3\pm0.6 \times 10^{10}$   \\
Hercules-A & --                    & 3.1$\pm1.1 \times 10^{40}$  & --  \\
Zw8193     & 1.5$\times 10^{42}$   & 1.1$\pm0.1 \times 10^{41}$  & $<4.3 \times 10^{10}$  \\
Zw8197     & 2.1$\times 10^{41}$   & $<1.8 \times 10^{40}$       & $1.1\pm0.3 \times 10^{10}$  \\
Zw8276     & 1.7$\times 10^{41}$   & 2.0$\pm0.2 \times 10^{40}$  & $8.2\pm1.2 \times 10^{9}$   \\
Cygnus-A   & (6.5$\times 10^{42}$) & 5.8$\pm0.2 \times 10^{40}*$  & ($<9.0\times 10^{9}$)  \\
A2390      & 9.8$\times 10^{41}$   & 1.2$\pm0.3 \times 10^{41}$  & $<4.9 \times 10^{10}$  \\
A2597      & (5.2$\times 10^{41}$) & 1.0$\pm0.1 \times 10^{41}$  & $8.1\pm3.3 \times 10^{9}$  \\
\end{tabular}
\caption{Table of 1-0 S(1) (or equivalent) molecular line luminosities
and CO-derived molecular gas masses. The 1-0 S(1) line luminosity is 
derived from either S(3) or S(5) if S(1) is not covered in the observed
spectral range. The CO-derived molecular gas masses in bracketts are ones
derived from the literature by Edge (2001). The S(1) line luminosities for
A478 and Cygnus-A
(asterisked) are derived from the line fluxes of Jaffe, Bremer \& van der Werf (2001)
and Wilman et al. (2000), respectively.}
\end{table*}

\subsection{Comparison to starbursts and LINERS}

% Murphy et al astro-ph/0010077
% Larkin et al astro-ph/9708097 ApJS 114, 59

The most comparable objects to the peculiar central cluster
galaxies studied here are starbursts as they are both 
known to harbour young stars, cold molecular gas and dust.
The infra-red spectroscopy of Ultraluminous Infra-red Galaxies
by Murphy et al. (1999) shows much stronger Pa$\alpha$ and
Br$\gamma$ and weaker H$_2$ lines compared to this study.
In addition the relatively strong [FeII] emission found
in this study is not observed in starbursts.
The excitation temperature of the H$_2$ lines
is similar to that found in this study and others
(see Wilman et al. 2002) but the properties of these two classes of objects
are significantly different in the near-infrared.

From the properties of the optical emission lines,
central cluster galaxies share many properites with
LINERs. Infra-red spectroscopy of LINERs by
Larkin et al.\ (1998) finds strong molecular lines and
a small fraction with high ionization lines so the
resemblance follows into the infra-red as noted by 
Jaffe et al.\ (2001). This
comparison also applies to the [FeII] strength as
LINERs (unlike starbursts) strong in [FeII] and [OI] and
weak in Br$\gamma$ (Larkin et al.\ 1998). 
The question of whether the underlying excitation mechanism 
behind both classes of object is the same is an open one.
The bulk of the literature on LINERs favours X-ray
heating and/or shocks and this is discussed further in
Wilman et al.\ (2002).

\subsection{CO absorption lines}

In a few of the lower redshift galaxies our spectra cover the 
CO stellar absorption bandhead at 2.3$\mu$m. The strength
of this feature should provide constraints of the relative
mass distribution on the underlying population of stars
in the galaxies. Using the definition of the CO
equivalent width of James \& Mobasher (2000), we calculate values 
of  3.57, 3.27, 3.96, 3.21, 3.33, 3.45 and 2.89nm, with errors of 0.30nm, for  
RXJ0338+096, Hydra-A, A85, A1795, A2029, A2052 and A2199 respectively
(only the latter object is in the James \& Mobasher (2000)
sample at 3.23$\pm$0.25nm so consistent within the errors). 
The mean equivalent width of CO
of our seven values is 3.38nm compared to 3.35nm for 
the brightest cluster galaxies in James \& Mobasher (2000)
so our results are consistent despite the redshift distribution and 
resolution of our spectra  not being well suited to studying these lines.

\subsection{Spectral slope at 2$\mu$m}

The K-band spectra obtained in ths study are dominated by
the old stellar population. However, in the case of NGC1275
there is a strong contribution from the active nucleus
which shows a significant upturn at 2$\mu$m. This is in
part due to the presence of hot dust (Krabbe et al.\ 2000).
To test for any possible non-stellar contribution to 
our spectra we have determined the spectral slope
from 1.9 to 2.1$\mu$m (rest) for all spectra (excluding
emission lines). The results are presented in Figure 8
plotted against redshift.
Only one other system (A1068) appears to be significantly
different from the overall distribution (mean of 2.59 and
an rms of 0.30 excluding NGC1275, A1068 and all the $z>0.2$ objects
which have weak continuum). Unlike NGC1275, A1068
does not show any broad emission lines but does show
evidence for a strongly ionizing source ([OIII], [SiVI] and Br~{$\gamma$})
so this source may host a highly obscured, radio-quiet Seyfert nucleus.
Therefore, while the contribution from non-stellar sources (nuclear 
continuum and hot (T$>>100$~K) dust) is small in the vast majority 
of our spectra, the spectral slope around 2$\mu$m is a useful diagnostic
to a possible nuclear contribution. 

\begin{figure*}
\centerline{\psfig{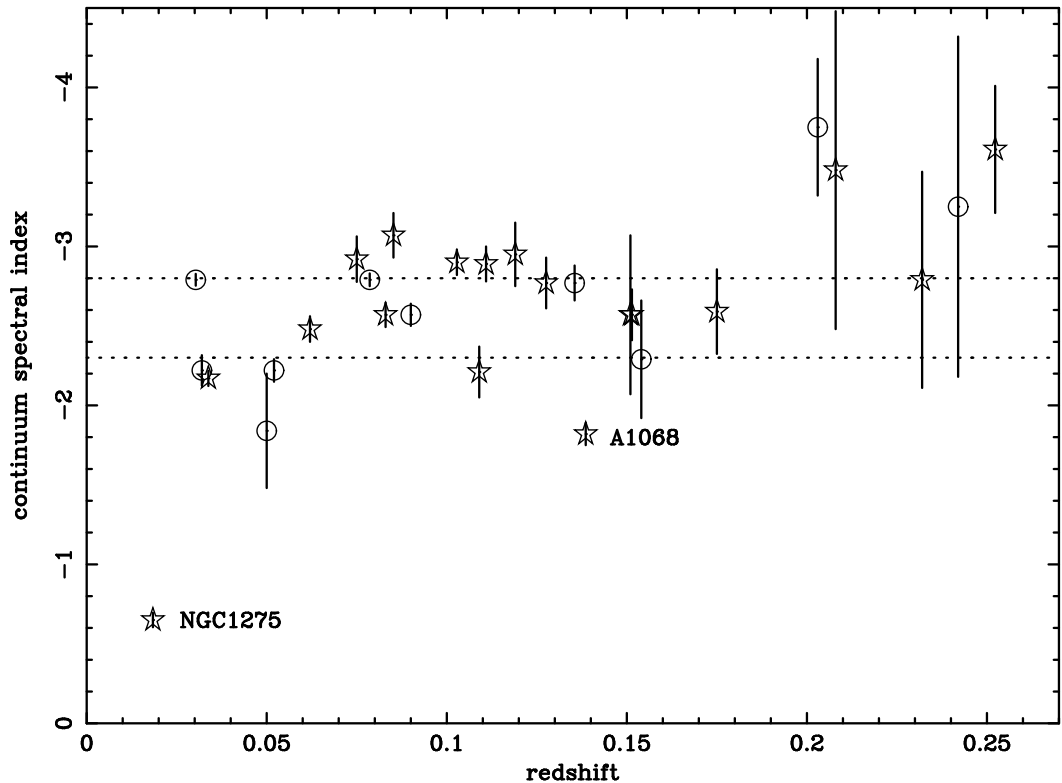}}
\caption{The spectral index of the continuum in the K-band between 1.9 and 2.1$\mu$m (rest)
for all objects. The stars mark systems with significant NIR line emission. The
dotted lines mark the 30 and 70 percentiles of the objects with z$<0.2$.}

\end{figure*}

\section{CONCLUSIONS}

The study presented in this paper confirms that there
is a remarkably good correspondance between the warm and cool
molecular gas and  ionized hydrogen
in intensity for central galaxies in 
cooling flows. These components appear to be common to this
class of objects and indicate that substantial masses of 
irradiated molecular gas lie in the centres of clusters.
What is required now is to establish the spatial correspondance
between the observed components to these systems through
narrow-band imaging and intergral field spectroscopy.

In the companion paper (Wilman et al. 2002) we present an analysis
of the likely excitation mechanisms for the molecular lines observed.

The ultimate question of whether the total mass of cold molecular clouds
deposited by the cooling flow
is comparable to that predicted from X-ray observations cannot
be addressed until observations of the far-infrared lines
that should be emitted by the coolest 
clouds as they themselves cool. These lines are within reach of
{\it SOFIA, SIRTF} and {\it FIRST} so this issue may be resolved in 
the next 2--4 years.

\section*{ACKNOWLEDGEMENTS}

UKIRT is operated by the Joint Astronomy Centre on behalf of the United Kingdom
Particle Physics and Astronomy Research Council. 
RJW acknowledges support from the EU Marie Curie Fellowship Scheme,
RJW and RMJ acknowledge support from PPARC, 
and ACE, CSC, ACF and SWA thank the Royal Society for support.
The analysis in this paper uses the ASURV Rev 1.2 package (LaValley, Isobe \& Feigelson 1992).

\end{document}